# Unravelling Distance-Dependent Inter-Site Interactions and Magnetic Transition Effects of Heteronuclear Single Atom Catalysts on Electrochemical Oxygen Reduction

Tong Yang[a], Jun Zhou[b], Xiaoyang Ma[c], Keda Ding[a], Kay Chen Tan[d], Ge Wang[e], Haitao Huang[a], and Ming Yang [a, *]

[a] Department of Applied Physics, The Hong Kong Polytechnic University, Hung Hom, Hong Kong SAR, China

[b] Institute of Materials Research & Engineering, A*STAR (Agency for Science, Technology and Research), 2 Fusionopolis Way, Innovis, Singapore 138634, Singapore

[c] School of Information Science and Engineering, Shandong University, 72 Binhai Road, Qingdao 266237, China

[d] Department of Computing, The Hong Kong Polytechnic University, Hung Hom, Hong Kong SAR, China

[e] Beijing Advanced Innovation Center for Materials Genome Engineering, Beijing Key Laboratory of Function Materials for Molecule & Structure Construction, School of Materials Science and Engineering, University of Science and Technology Beijing, Beijing 100083, China

[*] To whom the correspondence should be addressed: mingyang@polyu.edu.hk (M. Y.)

**Abstract**

Inter-site interactions between single atom catalysts (SACs) in the high loading regime are critical to tuning the catalytic performance. However, the understanding on such interactions and their distance dependent effects remains elusive, especially for the heteronuclear SACs. In this study, we reveal the effects of the distance-dependent inter-site interaction on the catalytic performance of SACs. Using the density functional theory calculations, we systematically investigate the heteronuclear iron and cobalt single atoms co-supported on the nitrogen-doped graphene ($FeN_4$-C and $CoN_4$-C) for oxygen reduction reaction (ORR). We find that as the distance between Fe and Co SACs decreases, $FeN_4$-C exhibits a reduced catalytic activity, which can be mitigated by the presence of an axial hydroxyl ligand, whereas the activity of $CoN_4$-C shows a volcano-like evolution with the optimum reached at the intermediate distance. We further unravel that the transition towards the high-spin state upon adsorption of ORR intermediate adsorbates is responsible for the decreased activity of both $FeN_4$-C and $CoN_4$-C at short inter-site distance. Such high-spin state transition is also found to significantly shift the linear relation between hydroxyl (*OH) and hydroperoxyl (*OOH) adsorbates. These findings not only shed light on the SAC-specific effect of the distance-dependent inter-site interaction between heteronuclear SACs, but also pave a way towards shifting the long-standing linear relations observed in multiple-electron chemical reactions.

## Introduction

As a bridge between homogeneous and heterogeneous catalysts, single atom catalysts (SACs) have received tremendous attention due to the maximum atom utilization, strong quantum confinement effect and steric hindrance effect.[1-3] Over the past decade, SACs have been reported to exhibit high catalytic performance towards various electrochemical reactions, such as hydrogen evolution reaction (HER),[4-9] oxygen evolution/reduction reaction (OER/ORR),[10-14] nitrogen reduction reaction (NRR)[15-18] and carbon dioxide reduction reaction ($CO_2RR$).[19, 20] Amounts of efforts have also been put in engineering the local coordination environment of SACs, aiming to further fine-tune or modify their intrinsic catalytic performance.[20-28] For instance, four-fold carbon-embedded nitrogen coordinated iron and cobalt single atoms ($FeN_4$-C and $CoN_4$-C) are well-known for their high activity and selectivity towards the $4e^-$ ORR pathway to water ($H_2O$) and have been proposed as promising alternative electrocatalysts to benchmark platinum (Pt).[10, 11] The catalytic activity of $FeN_4$-C can be further enhanced upon selective cleavage of adjacent C-N bonds or introductions of phosphorous dopants.[21, 22] For $CoN_4$-C, either deliberate substitutions of the nitrogen atoms with oxygen dopants or the presence of oxygen adsorbates near the $CoN_4$ moiety could shift its catalytic selectivity towards the $2e^-$ ORR pathway to hydrogen peroxides ($H_2O_2$).[24, 25]

Increasing the loading and the associated real active site density is of great importance to the overall catalytic performance of SACs. Kucernak *et al.* reported that for $FeN_x$-C, only a few active Fe sites are exposed and accessible for ORR, whereas most sites are buried.[29] The concave-shaped $FeN_4$-C was later synthesized at the loading of 2.78 wt%, where both the external surface area and exposed active Fe sites are maximized. Consequently, the as-synthesized $FeN_4$-C could achieve the US Department of Energy 2018 target set for platinum-group metal-free catalysts in proton exchange membrane fuel cells.[30] On the other hand, with the increase in the SAC loading, the interaction between active sites gradually comes into play. Based on density functional theory (DFT) calculations, Li *et al.* observed that carbon monoxide (CO) adsorption could induce a unique oscillatory long-range spin coupling between adjacent Fe atoms in the three-fold $FeN_3$-C when their distance is reduced to about 11 Å.[31] Han *et al.* predicted that the inter-site Fe distance also has a strong impact on the ORR performance of $FeN_3$-C.[32] More recently, the inter-site interaction between Fe sites in $FeN_4$-C has been comprehensively investigated for ORR applications.[33] A series of $FeN_4$-C were experimentally synthesized, where the inter-site Fe distance can be controlled in a range from several nanometres to subnanometres. It was found that the intrinsic ORR activity starts to increase when the inter-site distance is less than about 12 Å and reaches the optimum when Fe sites are around 7 Å apart. These studies manifest the significant role of the inter-site interaction in the catalytic performance of homonuclear SACs as loadings increase. Engineering the atomic species of neighbouring single atoms and forming heteronuclear SACs could be another approach to tune their performance, in which the inter-site interaction between heteronuclear single atoms might be even more pronounced when they are approaching one another. Indeed, various heteronuclear dual atom catalysts, where single atoms of different species are either bounded to or isolated from one another, have been synthesized in experiments and inter-site interactions have also been implied by the observed variation of the catalytic performance.[34-39] Nevertheless, the distance dependence of the inter-site interaction between heteronuclear single atoms and the induced effects on their catalytic activities still remain underexplored. Noting the superior performance of $FeN_4$-C and $CoN_4$-C towards the 4$e^-$ ORR pathway,[10, 11] in this study, they were chosen as a model system of heteronuclear SACs to systematically investigate the distance-dependent inter-site interaction and their effects on the ORR catalytic performance.

## Results and Discussion

### Structural Models and Stability

To scrutinize the distance-dependent interaction, in this study, a pair of heteronuclear $TMN_4$ moieties (TM=Fe, Co) were accommodated by a series of graphene substrates, of which the supercell size ($n$) and the average inter-site distance ($\langle d_{site} \rangle$) range from 3×3 to 11×11 and from ~4 Å to ~16 Å, respectively (**Figure S1** in the Supplementary Information and **Figure 1**). One moiety ($TM_{cen}N_4$) is located near the centre of the model and the other ($TM_{cor}N_4$) near the corner. As shown in **Figures 1a-b** at $n = 5$, two kinds of configurations were considered with slightly different average inter-site distances (**Figure 1c**). In configuration I (II), the line connecting $TM_{cen}$ and $TM_{cor}$ is parallel to (slightly deviates from) the diagonal, which is hereafter denoted as $TM_{cen}$-$TM_{cor}$@$n×n$-I ($TM_{cen}$-$TM_{cor}$@$n×n$-II). The 9×9 graphene supercell with one $TMN_4$ moiety embedded, which has $<d_{site}>$ of 22.21 Å and is denoted as $TMN_4$@9×9-S, was here used to simulate the isolated $TMN_4$-C. It is noted that in $Fe_{cen}$-$Co_{cor}$@3×3-II the $Fe_{cen}N_4$ and $Co_{cor}N_4$ moieties share a nitrogen dopant, whereas they are isolated from each other in the other models considered (**Figure S1**). Due to the small inter-site distance in $Fe_{cen}$-$Co_{cor}$@3×3, they can also be thought of as dense unbonded dual atom catalyst systems. Uniform dispersed $FeN_4$-C with an average inter-site distance of $5 \pm 1$ Å has been reported experimentally.[33] Very recently, high-loading heteronuclear SACs with the inter-site distance down to subnanometers have also been demonstrated to be experimentally achievable.[40]

In comparison with conventional catalysts, SACs have lower stability because of the high surface free energy and the low coordination number.[41] Therefore, the stability of $Fe_{cen}$-$Co_{cor}$@$n×n$ was herein examined. First, the relative stability between the two configurations of $Fe_{cen}$-$Co_{cor}$@$n×n$ was compared. Note that the comparison was not made at $n = 3$ because of the exclusive co-sharing of a nitrogen dopant in $Fe_{cen}$-$Co_{cor}$@3×3-II (**Figure S1**). As shown in **Figure 1c**, configuration II is more energetically stable than configuration I across the concerned supercell size range. The energy difference between configuration I and II, $E(n × n - \mathrm{I}) - E(n × n - \mathrm{II})$, reaches the maximum at $n = 7$, above which it quickly decreases. Such a quick decrease somewhat indicates that the inter-site interaction between $Fe_{cen}$ and $Co_{cor}$ is very weak or negligible when the supercell size (the average inter-site distance) is above 11×11 (~16Å).

The thermodynamic and electrochemical stabilities of $Fe_{cen}$-$Co_{cor}$@$n×n$ can be further evaluated by the formation energy ($E_f$) and the dissolution potential ($U_d$) as follows:

$$E_f(\mathrm{TM}) = E(tot) - E(V_{\mathrm{TM}}) - E(\mathrm{TM}_{\mathrm{bulk}}) \quad (1)$$

$$U_d(\mathrm{TM}) = U_d^o(\mathrm{TM}_{\mathrm{bulk}}) - \frac{E_f(\mathrm{TM})}{Ne} \quad (2)$$

where $E(tot)$ and $E(V_{\mathrm{TM}})$ are the total energy of $Fe_{cen}$-$Co_{cor}$@$n×n$ without and with a TM vacancy ($V_{\mathrm{TM}}$; TM=$Fe_{cen}$ or $Co_{cor}$), respectively. $E(\mathrm{TM}_{\mathrm{bulk}})$ is the per-atom energy of the bulk-phase TM. $U_d^o(\mathrm{TM}_{\mathrm{bulk}})$ is the standard dissolution potential of the bulk-phase TM and $N$ the number of electrons which are involved in the dissolution. It has been reported that $E_f(\mathrm{TM}) < 0$ eV and $U_d(\mathrm{TM}) > 0$ V hold for most experimentally synthesized SACs.[42] Therefore, we applied this set of stability criteria, as summarized in **Figure 1d**. It can be seen that all $Fe_{cen}$-$Co_{cor}$@$n×n$-II considered in this work meet the criteria, thereby implying their stability against aggregation and dissolution. With the decreasing inter-site distance, the formation energy (the dissolution potential) of either $Fe_{cen}$ or $Co_{cor}$ increases (decreases), suggesting a lower stability when they are approaching one another. This is in line with the experimental observations that syntheses of SACs with high loadings remain a major challenge.[3] In addition, the lower formation energy of $Co_{cor}$ than that of $Fe_{cen}$ suggests a stronger binding of the former with the neighbouring nitrogen dopants, which is consistent with the shorter $Co_{cor}$-N bonds (see **Figure S2**). In **Figure 1d**, an exception was observed for $Fe_{cen}$ in $Fe_{cen}$-$Co_{cor}$@3×3-II, which is likely attributed to the nitrogen dopant co-sharing. Hereafter, we will focus on $Fe_{cen}$-$Co_{cor}$@$n×n$-II ($n = 9, 7, 5$ and $3$) and $Fe_{cen}$-$Co_{cor}$@3×3-I.

**Catalytic Activities of Completely Isolated FeN$_4$-C and CoN$_4$-C**

Prior to explorations of the distance-dependent inter-site interaction, we studied the catalytic activity of completely isolated FeN$_4$-C and CoN$_4$-C embedded in the 9×9 graphene supercell, *i.e.,* Fe@9×9-S and Co@9×9-S. Before the adsorption of an oxygen molecule (O$_2$), both the Fe and Co atoms lie in the graphene substrate plane. Due to the energetic preference for the end-on configuration, upon O$_2$ adsorption, they slightly protrude from the plane by 0.197 Å and 0.187 Å for the Fe and Co atoms, respectively. Our calculations show that the side-on O$_2$ adsorption is not stable on either the isolated Fe or Co site and would transform to the end-on configuration. The adsorption elongates the O-O bond length up to 1.284 Å on the Fe site and 1.274 Å on the Co site, implying that the Fe site can activate O$_2$ more significantly. As a result, the first hydrogenation to *OOH (* denotes an adsorbed state) is more exothermic on the Fe site than on the Co site, as displayed in **Figure 2**. According to the calculated Gibbs free energies, it was found that the isolated Co site also binds other ORR intermediate adsorbates slightly weakly compared with the isolated Fe site (**Figure 2b**). This might be ascribed to the relatively strong binding of the Co atom with the nitrogen-doped graphene substrate (see **Figure 1d**). **Figure 2** shows that the potential-determining step (PDS) is the first hydrogenation to *OOH for the isolated Co site, whereas it is the third hydrogenation step from *O to *OH for the isolated Fe site. As an effective thermodynamic descriptor, the calculated limiting potential ($U_L$) of 0.980 V for the Fe site and 0.925 V for the Co site implies their high catalytic activity towards ORR, between which the performance of Fe SACs is better. It is noteworthy that the superior ORR catalytic activities of Fe and Co SACs and their relative activity trend of Fe SACs > Co SACs have been well demonstrated in experiments.[10, 11, 37, 43, 44] These agreements suggest the reliability of our calculations.

**Distance-Dependent Catalytic Activities of Fe$_{cen}$ and Co$_{cor}$ in Fe$_{cen}$-Co$_{cor}$@$n$×$n$**

The dependence of the catalytic activity of Fe$_{cen}$ on the distance with its neighbouring Co$_{cor}$ was first studied. **Figure 3a** shows the impact of the inter-site distance on O$_2$ adsorption. When they are around 12.85 Å apart (9×9-II), O$_2$ adsorption on Fe$_{cen}$ is very similar to that at $\langle d_{\text{site}} \rangle =$ 22.21 Å (9×9-S). In this case, the side-on configuration of O$_2$ adsorption is found meta-stable with its energy being 0.088 eV higher than the end-on counterpart. As $\langle d_{\text{site}} \rangle$ decreases to 9.99 Å (7×7-II), however, the side-on configuration becomes energetically favourite. Interestingly, the energetic preference for the end-on configuration recovers with the further decrease in the inter-site distance (5×5-II at $\langle d_{\text{site}} \rangle = 7.14$ Å). When Fe$_{cen}$ and Co$_{cor}$ are as close as 4.27 Å

(3×3-I) or 4.37 Å (3×3-II), the side-on configuration is energetically favoured again. As a result, an alternating preference for the side-on and end-on configurations is observed for $O_2$ adsorption on $Fe_{cen}$ (**Figure 3a**). The O-O bond (Fe-O bonds) in the side-on configuration is found longer (shorter) than in the end-on configuration, indicative of a more efficient $O_2$ activation in the side-on configuration.

The evolution of Gibbs free energy along the $4e^-$ ORR pathway has been calculated for $Fe_{cen}$ in $Fe_{cen}$-$Co_{cor}$@$n\times n$ and visualized in **Figures 3b-f**. For clarity, $TM_{cen}$ and $TM_{cor}$ of $TM_{cen}$-$TM_{cor}$@$n\times n$ are hereafter denoted as $TM_{cen}$@$n\times n$ and $TM_{cor}$@$n\times n$, respectively. It is noted that adsorption of *OOH has a lower Gibbs free energy at $n = 7$ and 3 than at $n = 9$ and 5, consistent with the more efficient $O_2$ activation in the side-on configuration (**Figure 3a**). For $Fe_{cen}$@9×9-II, our calculations show that the presence of $Co_{cor}$ slightly modulates the energetics of ORR intermediate adsorbates and PDS is shifted from the hydrogenation of *O to *OH for Fe@9×9-S to that of *OH to $H_2O$ (see **Figures 2a** and **3b**). Notwithstanding, the calculated limiting potential of $Fe_{cen}$@9×9-II ($U_L = 0.990$ V) is very similar to that of Fe@9×9-S ($U_L = 0.980$ V), which indicates negligible change in the catalytic performance towards ORR. As the inter-site distance decreases to 9.99 Å (7×7-II), however, the binding strength of *OH is excessively increased by 0.545 eV (**Figure 3c**). With respect to $Fe_{cen}$@9×9-II, PDS remains unchanged for $Fe_{cen}$@7×7-II but the dramatically strengthened *OH adsorption leads to a significant decrease in the limiting potential from 0.990 V to 0.445 V and thus a substantial decrease in the ORR catalytic activity. **Figures 3d-f** show that the adsorption of *OH turns out to be even stronger as $Fe_{cen}$ and $Co_{cor}$ are further approaching one another. This results in further catalytic performance deterioration of $Fe_{cen}$@$n\times n$.

It is worth noting that for $Fe_{cen}$-$Co_{cor}$@3×3-II $Fe_{cen}$ and $Co_{cor}$ are as close as 3.21 Å along the nitrogen co-sharing direction. In this case, $O_2$ and *OOH are found to energetically prefer a bridge-like side-on adsorption configuration, where a $Fe_{cen}$-O-O-$Co_{cor}$ chain is formed. The O-O bond is cleaved by reducing *OOH to two *OH with one on $Fe_{cen}$ and the other on $Co_{cor}$. From **Figure 3e**, it is clear that $Fe_{cen}$@3×3-II can activate and cleave the O-O bond the most efficiently with the assistance of neighbouring $Co_{cor}$. It has been recently reported that the adjacent Mn-W pair could efficiently boost the polarization/activation of the N-N triple bond of $N_2$ for nitrogen fixation.[45] Despite the efficient $O_2$ activation, the excessively strong adsorption of *OH limits the catalytic performance of $Fe_{cen}$@3×3-II towards ORR.

In contrast to $Fe_{cen}@n×n$, $Co_{cor}@n×n$ always energetically favours the end-on configuration of $O_2$ adsorption across the considered supercell size (inter-site distance) range (**Figure 4a**). As the distance between $Fe_{cen}$ and $Co_{cor}$ decreases, the O-O bond length first increases, reaches the maximum for $Co_{cor}@5×5$-II ($\langle d_{site} \rangle = 7.14$ Å) and then decreases. We noticed that the calculated limiting potential generally follows the same trend as the O-O bond length elongation, as illustrated in **Figures 4b-f**. $Co_{cor}@5×5$-II exhibits an optimal limiting potential of 1.022 V, implying the catalytic performance enhancement compared to the isolated Co@9×9-S ($U_L = 0.925$ V). It also surpasses the isolated Fe@9×9-S ($U_L = 0.980$ V) and $Fe_{cen}@9×9$-II ($U_L = 0.990$ V), and thus possesses the best ORR catalytic activity among the considered $Fe_{cen}$-$Co_{cor}@n×n$. This is in stark contrast to $Fe_{cen}@n×n$, in which the presence of neighbouring $Co_{cor}$ turns out to be detrimental to the ORR activity of $Fe_{cen}$, as discussed above. With regard to PDS of $Co_{cor}@n×n$, it remains the hydrogenation of $O_2$ to *OOH until the neighbouring $Fe_{cen}$ is as close as 4.27 Å (3×3-I) or 4.37 Å (3×3-II), where PDS is changed to the hydrogenation of *OH to $H_2O$ (see **Figures 4e-f**). It should be pointed out that in addition to the intrinsic catalytic activity of SACs that have been studied here, the active site density is another key factor in determining the overall catalytic performance of SACs. The reduced intrinsic activity of both $Fe_{cen}$ and $Co_{cor}$ at the short inter-site distance might be compensated by the greatly increased density of active sites.

**Distance-Dependent Spin States of $Fe_{cen}$ and $Co_{cor}$ in $Fe_{cen}$-$Co_{cor}@n×n$**

Previous investigations have suggested that for Fe- or Co-containing catalysts, their spin state is likely to be one of the dominant factors in determining the catalytic performance.[46-51] Therefore, we further analysed the spin state of $Fe_{cen}$ and $Co_{cor}$ in $Fe_{cen}$-$Co_{cor}@n×n$, in particular when the potential-determining ORR intermediate adsorbate is present. Before the ORR process proceeds, our calculations show that $Fe_{cen}@n×n$ and $Co_{cor}@n×n$ have a nearly constant magnetic moment of around 2 $\mu_B$ and 1 $\mu_B$, respectively, regardless of the inter-site distance. The calculated magnetic moments are almost the same as those of the Fe and Co atoms in Fe@9×9-S and Co@9×9-S, respectively. As for $Fe_{cen}@n×n$, since *OH is the potential-determining intermediate adsorbate across $n$ (see **Figure 3**), the spin state of $Fe_{cen}$ with *OH adsorbed was studied and summarized in **Figure 5a**. It can be seen that when *OH is adsorbed on $Fe_{cen}$, its magnetic moment exhibits a step-like transition with the decrease in the inter-site distance. $Fe_{cen}$ shows a low-spin state in 9×9-II and transits to a high-spin state when the distance reduces to 9.99 Å (7×7-II) and onwards. Interestingly, an analogous step-like behaviour is also observed for the Gibbs free energy of *OH ($G(*OH)$) – the sudden switch of

Fe$_{cen}$ from the low-spin to high-spin state in 7×7-II accompanies a sudden adsorption enhancement of *OH. For Co$_{cor}$, upon adsorption of *OH (potential-determining in 3×3-I and 3×3-II) and *OOH (potential-determining in other cases), it first remains the low-spin state and then changes to an intermediate spin state in 5×5-II and $n$×$n$-II at $n = 9, 7$ and 5, respectively (see **Figure 5b**). The further decrease in the inter-site distance (3×3-I and 3×3-II) turns Co$_{cor}$ into the high-spin state. Similarly, the adsorption strength of *OH and *OOH is also found to be positively correlated with the inter-site distance-dependent magnetic moment of Co$_{cor}$.

Despite being almost insensitive to the inter-site distance in the absence of adsorbates, the magnetic moment of Fe$_{cen}$ and Co$_{cor}$ shows a strong dependence when *OH or *OOH is adsorbed. Such an adsorption-induced distance dependence of the spin state of Fe$_{cen}$ and Co$_{cor}$ is likely linked to the adsorption-induced modulation of their coordination environment that is expected dependent on the inter-site distance owing to the distance-dependent binding strength of Fe$_{cen}$ and Co$_{cor}$ with the neighbouring N dopants (**Figure 1d**). In the absence of adsorbates, Fe$_{cen}$ and Co$_{cor}$ are found to always lie within the graphene substrate plane, which may be responsible for the distance-independent spin state. In the presence of adsorbates, however, such independence is broken, as manifested by the varied TM-O bond length ($d_{\text{TM-O}}$) and the protrusion height ($h_{\text{TM}}$) of TM (TM=Fe$_{cen}$ or Co$_{cor}$) in **Figure 5c**. For adsorption of *OH on Fe$_{cen}$, the TM-O bond length varies slightly, whereas Fe$_{cen}$ starts to significantly protrude from the substrate plane in 7×7-II. This implies that when Fe$_{cen}$ and Co$_{cor}$ are close to one another, adsorption of *OH could lead to a transition of Fe$_{cen}$ from a nearly in-plane coordination environment to the tetrahedral one. In addition, we found that this coordination environment transition occurs at the same inter-site distance (7×7-II) as the high-spin state transition. Such a simultaneous transition towards both the tetrahedral coordination and the high-spin state has been reported when a nitrogen atom acts as an axial ligand to attach to FeN$_4$-C, forming N-FeN$_4$-C.[50] It is worth noting that N-FeN$_4$-C has also been identified to be highly active towards ORR.[49, 50] Thus, we also evaluated the ORR performance of Fe$_{cen}$ in $n \times n$ at $n = 7, 5$ and 3 with an axial hydroxyl ligand attached, which is denoted as Fe$_{cen}^{*OH}$ (**Figure S3**). The calculated limiting potential of Fe$_{cen}^{*OH}$ ($U_L^{*OH}$) was displayed in **Figure 5a**. While the hydroxyl ligand significantly improves the catalytic activity towards ORR, it is still inferior to that of the far-separated Fe sites in 9×9-II and 9×9-S.

Similarly, when *OH or *OOH is adsorbed, Co$_{cor}$ also undergoes a transition towards the tetrahedral coordination and the high-spin state with the decrease in the inter-site distance (**Figures 5b-c**). In contrast to Fe$_{cen}$, this transition occurs at a much smaller inter-site distance

(3×3-I and 3×3-II). On the other hand, **Figure 5c** shows that the unique intermediate spin-state of Co$_{cor}$ can be associated with the elongation of the Co$_{cor}$-O bond before the ultimate transition to the high-spin state. These observations indicate a stronger binding strength of Co$_{cor}$ than that of Fe$_{cen}$ at the same inter-site distance, in agreement with the lower formation energies of the former (**Figure 1d**).

**Distance-Dependent Electronic Structures of Fe$_{cen}$ and Co$_{cor}$ in Fe$_{cen}$-Co$_{cor}$@$n$×$n$**

The electronic structure of Fe$_{cen}$ and Co$_{cor}$ was also investigated in order to gain a deeper insight into the adsorption-induced spin state transition. Fe$_{cen}$-Co$_{cor}$@9×9-S, Fe$_{cen}$-Co$_{cor}$@5×5-II and Fe$_{cen}$-Co$_{cor}$@3×3-I were chosen as representative systems for the large, intermediate and small inter-site distance, respectively. Wherein, Co$_{cor}$ is in the low-, intermediate- and high-spin state, respectively, upon adsorption of *OH or *OOH (see **Figure 5b**). **Figure S4** illustrates the projected density of states (PDOS) onto the Fe site without adsorbates attached. As the distance of neighbouring Co$_{cor}$ is reduced, the electronic structure of Fe$_{cen}$ is affected to some extent, especially the Fe-$d_{x^2-y^2}$ orbital which is significantly broadened in 3×3-I due to the hybridization with the neighbouring Co-$d_{x^2-y^2}$ orbital. Nevertheless, the origin of the distance-independent magnetic moment of 2 $\mu_B$ remains unchanged – the nearly empty Fe-$d_{z^2}$ orbital and the partially occupied Fe-$d_{xz}$ and $d_{yz}$ orbitals in the spin-down channel (**Figure S4**). Upon adsorption of potential-determining *OH on Fe@9×9-S, the spin-up Fe-$d_{z^2}$ orbital couples to the molecular state of *OH and forms partially unoccupied anti-bonding state in the spin-up channel, thereby giving rise to the reduced magnetic moment of the Fe site in 9×9-S (right panel of **Figure 6**). As Fe$_{cen}$ and Co$_{cor}$ move to the intermediate inter-site distance (5×5-II), the significant adsorption-induced protrusion causes a large spin splitting for all Fe-$d$ orbitals, leading to more (less) unoccupied spin-down (spin-up) states and thus the high-spin state (middle panel of **Figure 6**). With the further decrease in the inter-site distance (3×3-I), the spin polarization of the Fe-$d_{x^2-y^2}$ and $d_{xy}$ turn out to decrease, making them partially re-occupied in the spin-down channel and thus a slightly reduced high-spin state (**Figure 5a** and left panel of **Figure 6**).

As for the pristine Co@9×9-S, the calculated PDOS in **Figure S5** shows a very similar electronic structure of the Co site to that of the Fe site in Fe@9×9-S, except that the one more valence electron of Co fully fills the Co-$d_{xz}$ and $d_{yz}$ orbitals in the spin-down channel. Thus, only the spin-down Co-$d_{z^2}$ orbital is empty, resulting in a magnetic moment of 1 $\mu_B$. This also holds for Co$_{cor}$ in Fe$_{cen}$-Co$_{cor}$@5×5-II (**Figure S5**). When Fe$_{cen}$ and Co$_{cor}$ further get closer

(3×3-I), however, the Co-$d_{z^2}$ orbital is fully occupied. Instead, the small inter-site distance enables the Co-$d_{xz}$ and $d_{yz}$ orbitals to resonate with the partially spin-polarized counterparts from Fe$_{cen}$ and hence be partially polarized as well, leading to the 1-$\mu_B$ magnetic moment of Co$_{cor}$ in Fe$_{cen}$-Co$_{cor}$@3×3-I (see **Figures S4-S5**). With potential-determining *OOH adsorbed on Co@9×9-S, the hybridization with the molecular states of *OOH reduces the spin polarization of the Co-$d_{z^2}$ orbital with the spin-up (spin-down) channel partially unoccupied (occupied), giving the low-spin state of Co (right panel of **Figure 7**). By comparison, the elongation of the Co-O bond in Fe$_{cen}$-Co$_{cor}$@5×5-II partially enhances the spin splitting of the Co-$d_{z^2}$ orbital, giving rise to the adsorption-induced intermediate-spin state of Co (middle panel of **Figure 7**). As for the adsorption-induced high spin state of Co$_{cor}$ in Fe$_{cen}$-Co$_{cor}$@3×3-I, **Figure 7** shows that it mainly originates from the large spin splitting of the Co-$d_{x^2-y^2}$ orbital as well as the partially unoccupied Co-$d_{xz}$, $d_{yz}$ and $d_{z^2}$ orbitals. Regarding adsorption of *OH, the electronic structure of Co was found to evolve in an analogous way with the inter-site distance, as shown in **Figure S6**. It is worth noting that compared with adsorption of *OOH, *OH interacts more strongly with the Co atom in 9×9-S, which pushes the originally occupied spin-up Co-$d_{z^2}$ orbital to above the Fermi level and hence leads to a complete spin depolarization of Co (**Figures 5b** and **S6**).

**Catalytic Activities of Co$_{cor}$ in Other TM$_{cen}$-Co$_{cor}$@5×5-II and Scaling Relations**

Among the sites considered above, Co$_{cor}$ in Fe$_{cen}$-Co$_{cor}$@5×5-II was found to be the most active site (**Figures 5a-b**). We further substituted Fe$_{cen}$ in this configuration with Mn, Co, Ni, Cu, Ir and Pt, and investigated the catalytic activity of Co$_{cor}$. In the single atom form, these elements have been reported experimentally synthesizable and stable during the ORR process.[23, 39, 52-54] The calculated Gibbs free energy diagrams are displayed in **Figure S7** and **Figure 8a** summarizes the associated limiting potential. One can see that Pt$_{cen}$ can further improves the catalytic activity of Co$_{cor}$, whereas it slightly decreases when Fe$_{cen}$ is replaced with Mn$_{cen}$, Co$_{cen}$ and Ir$_{cen}$. Compared with the inter-site distance, engineering the neighbouring single atom species shows a less extent to which the activity is modulated (**Figures 5b** and **Figure 8a**) at the intermediate inter-site distance.

It is well-known that the adsorption strength of *OH linearly scales with that of *OOH due to the binding similarity, which has imposed a harsh restriction on further optimization of the catalytic performance.[55, 56] This linear relation has also been recently found to hold for single atom catalyst systems.[57] With the distance-dependent inter-site interaction considered in the present study, we also explored in this regard. As shown in **Figure 8b**, there is a clear positive correlation between the Gibbs free energy of *OOH ($G(*OOH)$) and that of *OH ($G(*OH)$). Nevertheless, it was found that when the sites are divided into two groups, a better fitting result can be achieved for each group, as evidenced by the higher Pearson's r (see **Figures 8b** and **S8**). Intriguingly, this grouping well separates those sites for which adsorption of *OH or *OOH can induce the transition to the high-spin state from others. The former (latter) group corresponds to the sites in the light orange (olive) shaded region in **Figure 8b** and is hereafter denoted as group I (II). The isolated (9×9-S) and far-apart Fe and Co sites (9×9-II) fall into group II, where a similar linear relation to that reported previously is observed: $G(*OOH) = 1.071 \times G(*OH) + 2.759$ with the Pearson's r of 0.884 (the olive solid line in **Figure 8b**). As compared to group II, the sites in group I display a very different linear relation: $G(*OOH) = 0.451 \times G(*OH) + 3.461$ with the Pearson's r of 0.890 (the orange solid line in **Figure 8b**). We noted that dual mononuclear TMN$_4$-C has been predicted to significantly attenuate the linear relation between intermediate adsorbates of nitrogen reduction reaction (NRR) and thus leads to an enhanced NRR activity.[58] The bridge-on adsorption was there found to play a key role due to the small inter-site distance between mononuclear TM pairs. On the contrary, in the present study, the appreciable modification observed for the sites in group I just involves single-site adsorption. This may suggest the distance-dependent inter-site interaction or the adsorption-induced spin-state transition to be another potential approach to shift the linear relation, although it shifts in an unfavourable direction for the sites in group I (**Figure 9**).

## Conclusion

In summary, we systematically investigate the effects of the distance-dependent inter-site interaction on the stability, catalytic activity, magnetic and electronic properties of nitrogen-doped carbon supported heteronuclear iron and cobalt single atom catalysts in the high-loading regime (Fe$_{cen}$-Co$_{cor}$@$n \times n$). Fe$_{cen}$-Co$_{cor}$@$n \times n$ is predicted to be stable and experimentally synthesizable in the high loading regime. It is found that Fe$_{cen}$ and Co$_{cor}$ exhibit distinct distance-dependent catalytic activity towards ORR when their distance decreases. The catalytic activity of Fe$_{cen}$ is reduced but can be mitigated in the presence of an axial hydroxyl ligand. On the contrary, the catalytic activity of Co$_{cor}$ increases first, reaches the optimum at the intermediate inter-site distance, and then decrease at the short inter-site distance. The eventually reduced catalytic activity of both Fe$_{cen}$ and Co$_{cor}$ at short inter-site distance can be linked to the distance-dependent adsorption-induced transition towards the high-spin state.

Interestingly, such high-spin state transition could significantly shift the linear relation between *OH and *OOH. Our work not only suggests the experimental synthesis feasibility of dense heteronuclear iron and cobalt single atom catalysts, but also reveals the synergistic effects of high loading heteronuclear SACs on the intrinsic catalytic activity. These results might pave a way towards shifting the linear relation between intermediate adsorbates of ORR. It may also stimulate future explorations of this overlooked inter-site interaction between SACs as well as the associated SAC-specific distance-dependent effects for a broad spectrum of multi-electron chemical reactions.

## Conflicts of Interest

There are no conflicts to declare.

## Acknowledgements

This work was supported by the Hong Kong Polytechnic University (project ID: 1-BE47 and ZE2F). W. G. thanked National Natural Science Foundation of China (NSFC) for the financial support (No. 51972024). We acknowledged the Centre for Advanced 2D Materials and Graphene Research and the High-Performance Computing Centre at the National University of Singapore, and the National Supercomputing Centre of Singapore for providing the computing resources.

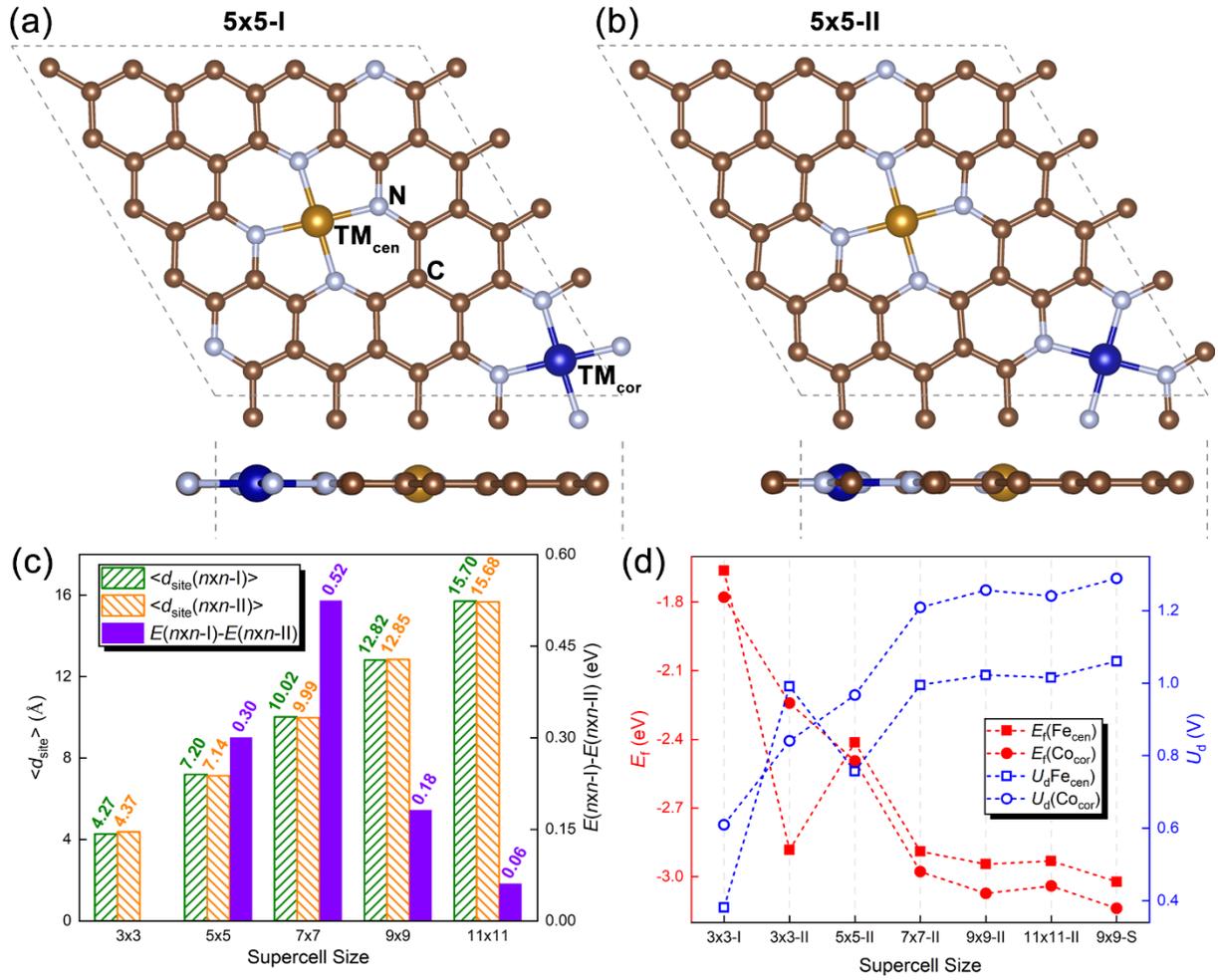

**Figure 1**. (a-b) The top (top row) and side view (bottom row) of (a) $TM_{cen}$-$TM_{cor}$@$n×n$-I and (b) $TM_{cen}$-$TM_{cor}$@$n×n$-II at $n = 5$. (c) The average distance ($\langle d_{site}\rangle$) between $Fe_{cen}$ and $Co_{cor}$ and the relative energetic stability between $Fe_{cen}$-$Co_{cor}$@$n×n$-I and $Fe_{cen}$-$Co_{cor}$@$n×n$-II as functions of the supercell size. $E(n×n\text{-I})$- $E(n×n\text{-II})$ is the energy difference between $Fe_{cen}$-$Co_{cor}$@$n×n$-I and $Fe_{cen}$-$Co_{cor}$@$n×n$-II. (d) The formation energy and the dissolution potential of $Fe_{cen}$ and $Co_{cor}$ as functions of the supercell size.

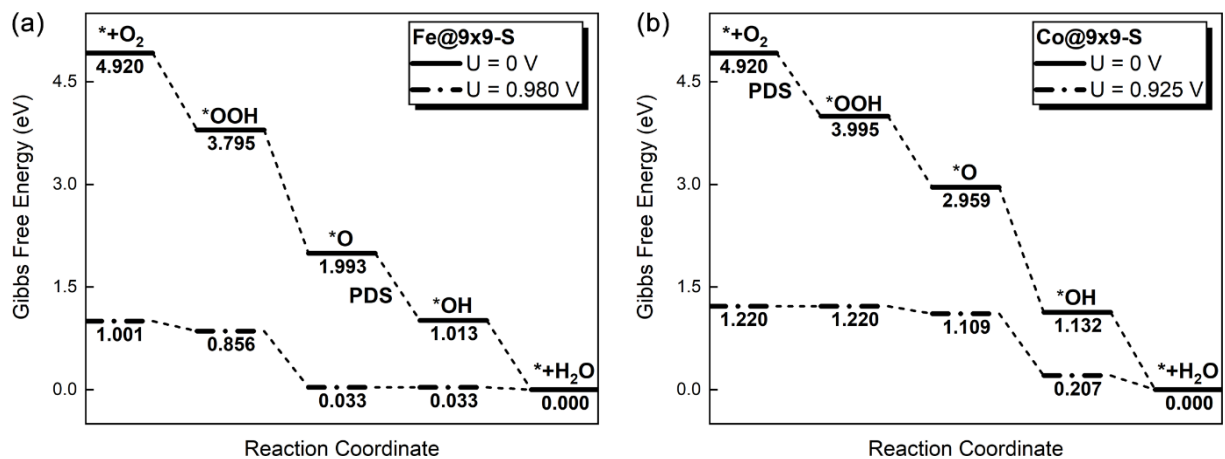

**Figure 2**. (a-b) The Gibbs free energy diagram for oxygen reduction reaction on (a) the Fe site of Fe@9×9-S and (b) the Co site of Co@9×9-S. The potential-determining step (PDS) is marked.

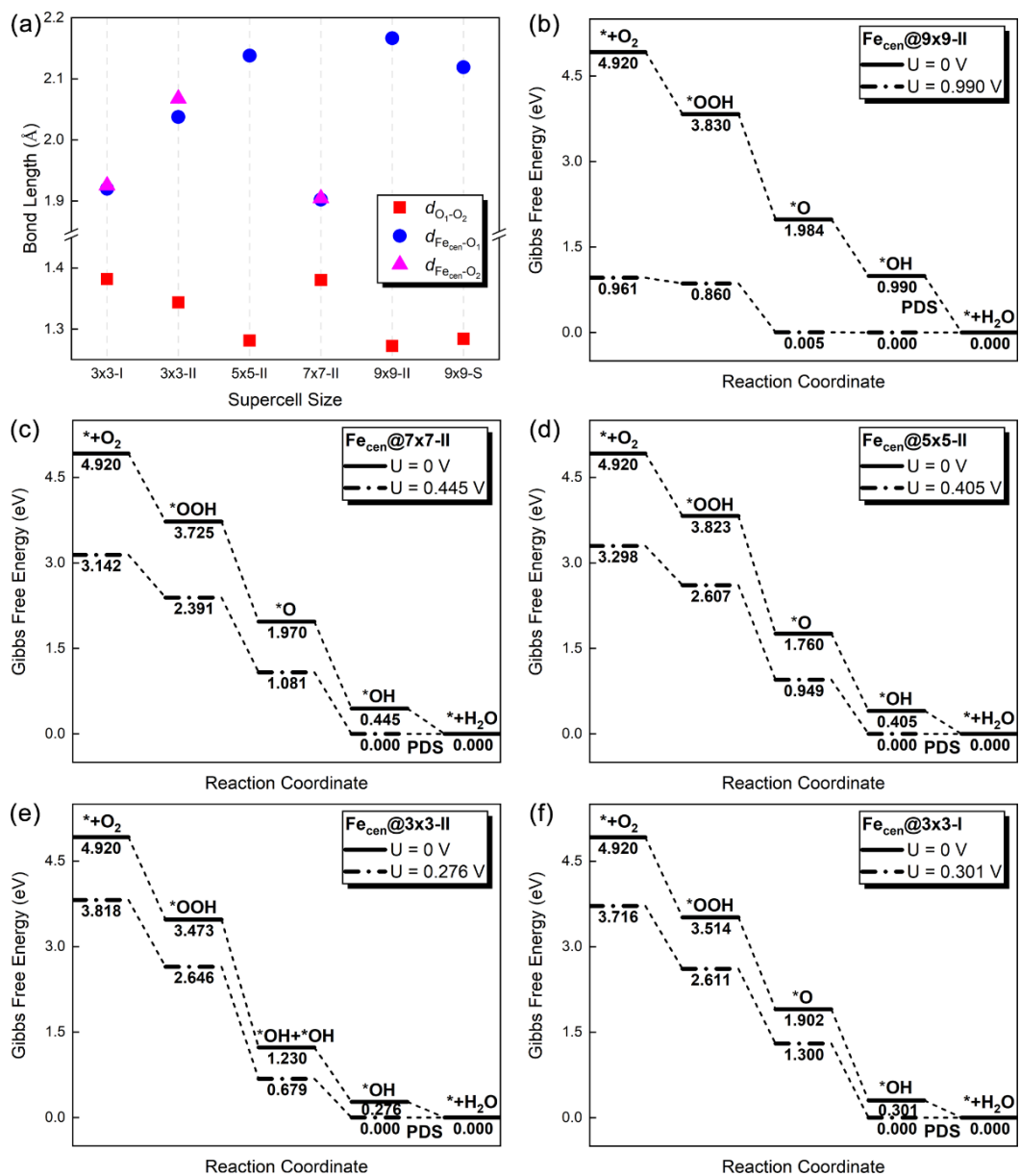

**Figure 3**. (a) The lengths of the O-O bond ($d_{O_1-O_2}$) and the bond(s) formed between O and $Fe_{cen}$ ($d_{Fe_{cen}-O_1}$ and $d_{Fe_{cen}-O_2}$) upon $O_2$ adsorption as functions of the supercell size. $d_{Fe_{cen}-O_2}$ is shown only for the side-on configuration of $O_2$ adsorption. (b-f) The Gibbs free energy diagram for oxygen reduction reaction on $Fe_{cen}$ in $Fe_{cen}$-$Co_{cor}$@$n\times n$-II at $n$ = (b) 9, (c) 7, (d) 5, (e) 3 and (f) $Fe_{cen}$-$Co_{cor}$@$3\times3$-I. The potential-determining step (PDS) is marked.

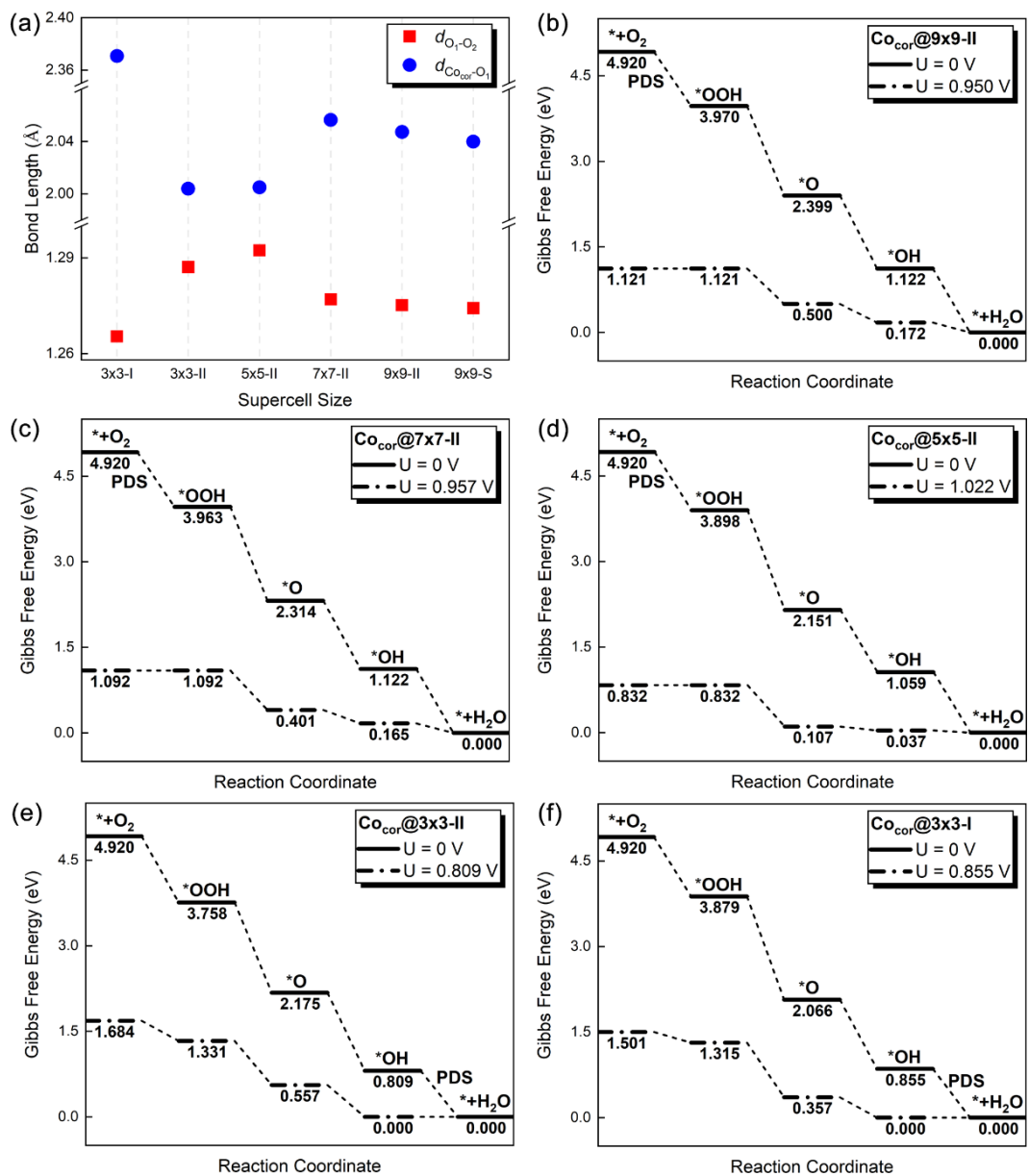

**Figure 4**. (a) The lengths of the O-O bond ($d_{O_1-O_2}$) and the bond formed between O and $Co_{cor}$ ($d_{Co_{cor}-O_1}$) upon $O_2$ adsorption as functions of the supercell size. (b-f) The Gibbs free energy diagram for oxygen reduction reaction on $Co_{cor}$ in $Fe_{cen}$-$Co_{cor}$@$n \times n$-II at $n =$ (b) 9, (c) 7, (d) 5, (e) 3 and (f) $Fe_{cen}$-$Co_{cor}$@3×3-I. The potential-determining step (PDS) is marked.

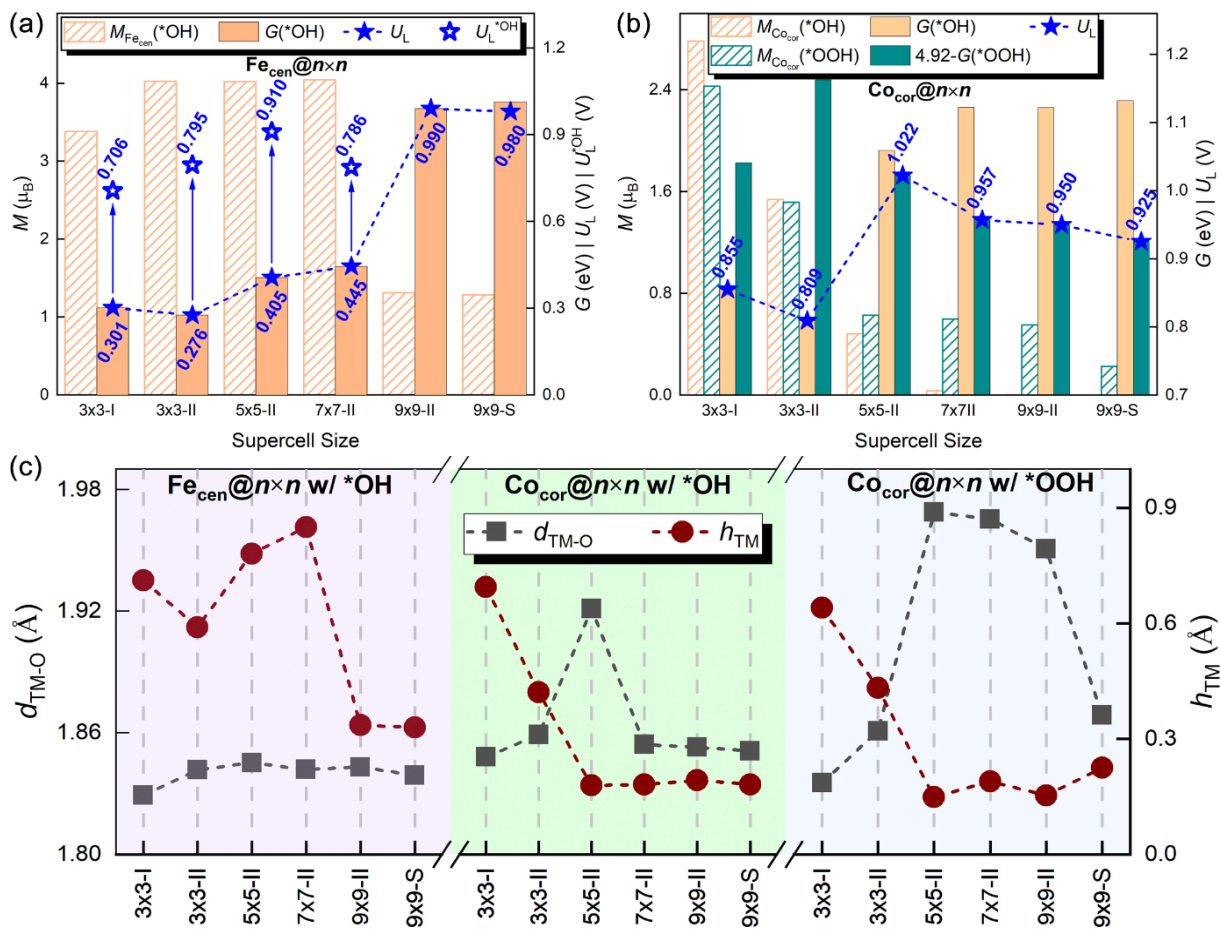

**Figure 5**. (a) The Gibbs free energy of *OH adsorption ($G(*OH)$), the ORR limiting potential ($U_L$) and the projected magnetic moment ($M$) on Fe$_{cen}$@$n \times n$. The limiting potential of Fe$_{cen}$ with an axial hydroxyl ligand attached ($U_L^{*OH}$) is marked by the empty blue star. (b) The Gibbs free energy of *OH and *OOH adsorption ($G(*OH)$ and $G(*OOH)$), the ORR limiting potential and the projected magnetic moment on Co$_{cor}$@$n \times n$. (c) The TM-O bond length ($d_{TM-O}$) and the protrusion height of TM ($h_{TM}$) in the presence of the *OH adsorbate on Fe$_{cen}$@$n \times n$ (TM=Fe$_{cen}$) and in the presence of the *OH and *OOH adsorbate on Co$_{cor}$@$n \times n$ (TM=Co$_{cor}$), respectively.

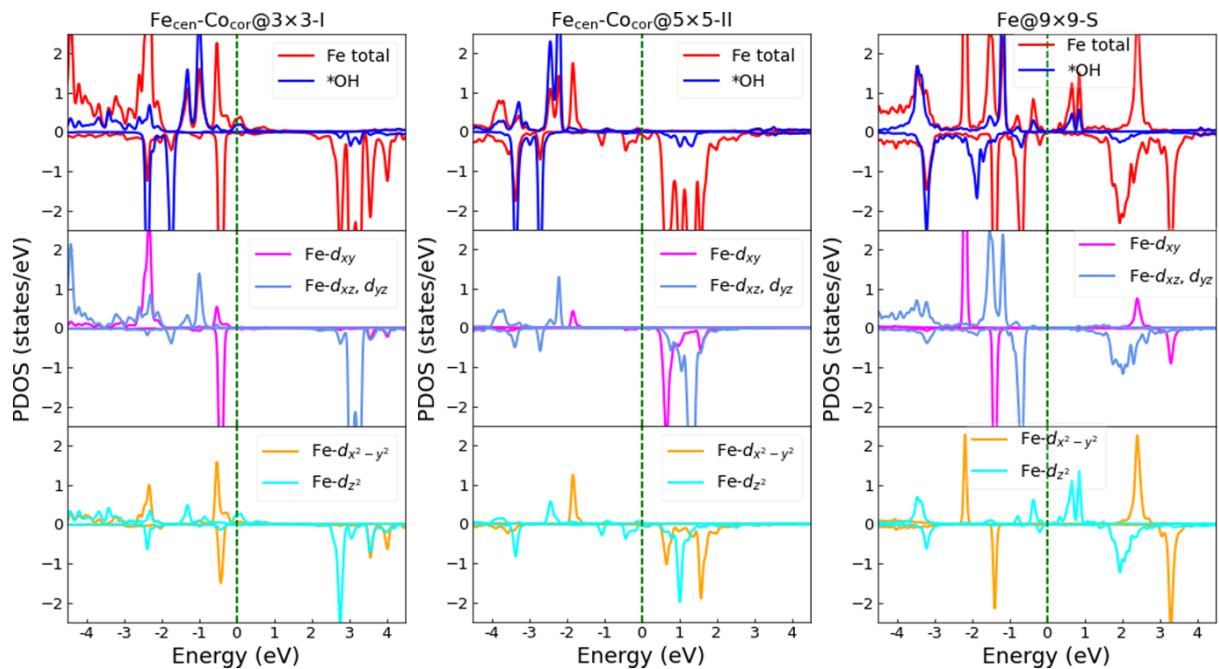

**Figure 6**. The projected density of states (PDOS) onto the Fe site and the hydroxyl adsorbate (*OH) of (left panel) Fe$_{cen}$-Co$_{cor}$@3×3-I, (middle panel) Fe$_{cen}$-Co$_{cor}$@5×5-II and (right panel) Fe@9×9-S when the *OH is adsorbed on to the Fe site. The Fermi level has been set to zero.

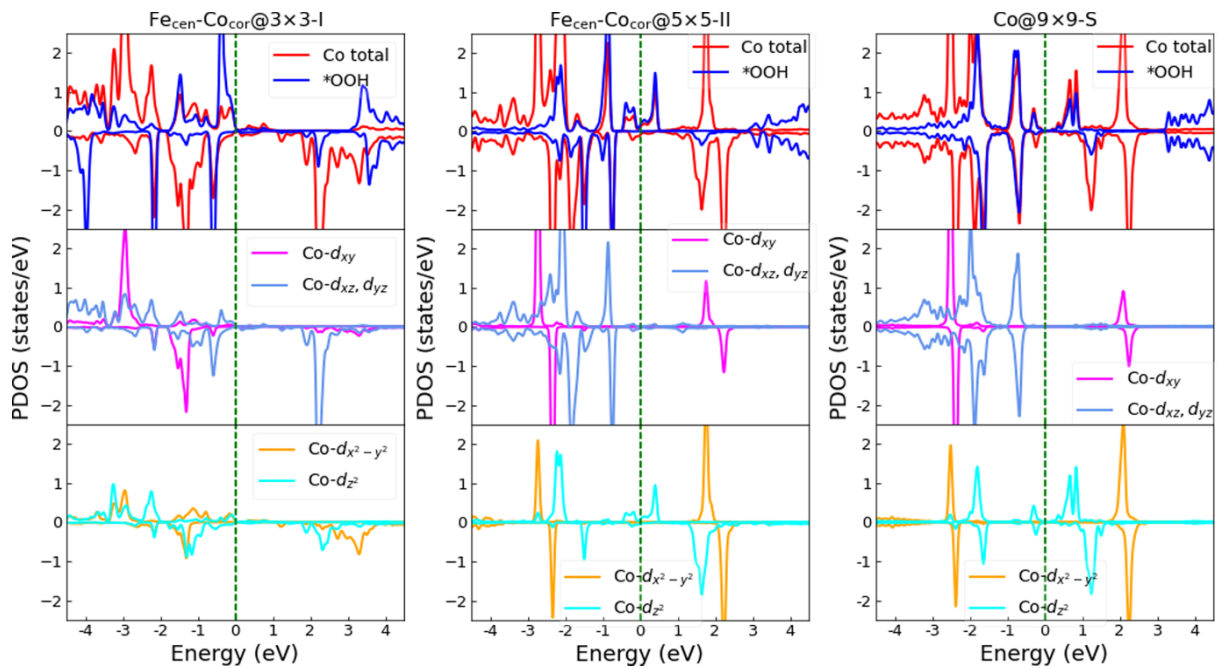

**Figure 7**. The projected density of states (PDOS) onto the Co site and hydroperoxyl (*OOH) of (left panel) Fe$_{cen}$-Co$_{cor}$@3×3-I, (middle panel) Fe$_{cen}$-Co$_{cor}$@5×5-II and (right panel) Fe@9×9-S when the *OOH is adsorbed on to the Co site. The Fermi level has been set to zero.

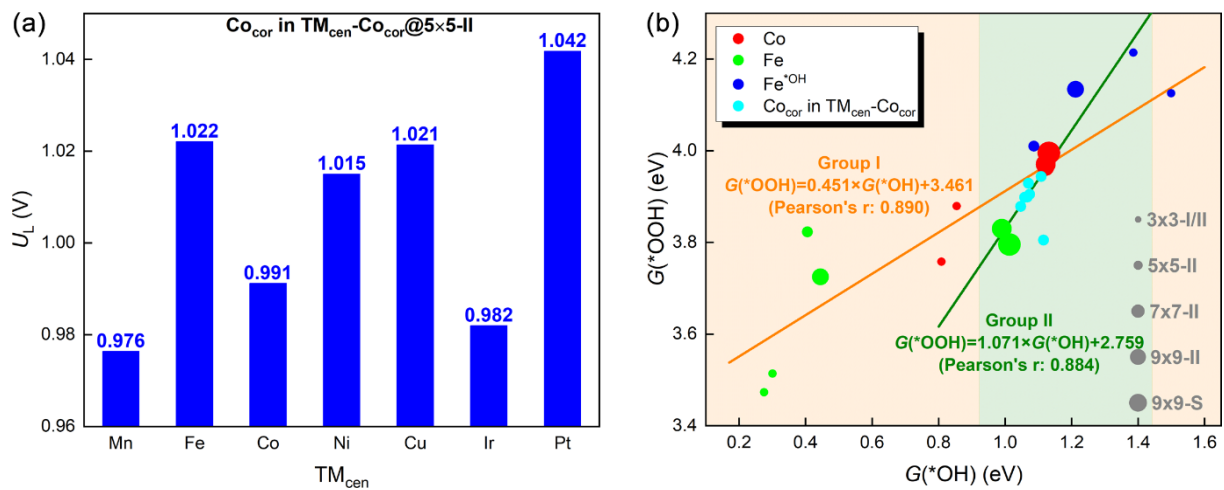

**Figure 8**. (a) The calculated limiting potential ($U_L$) for ORR on $Co_{cor}$ in $TM_{cen}$-$Co_{cor}$@5×5-II. (b). The Gibbs free energy of *OOH ($G(*OOH)$) versus that of *OH ($G(*OH)$) on all Co and Fe sites in $Fe_{cen}$-$Co_{cor}$@n×n-I/II, $TM_{cen}$-$Co_{cor}$@5×5-II, Fe@9×9-S, Co@9×9-S. $Fe^{*OH}$ denotes the Fe site with an axial hydroxyl ligand attached. The scatter size denotes the corresponding supercell size and type. The orange and olive solid lines are fitted to the points in the light orange (group I) and olive (group II) shaded regions, respectively, with the Pearson's r marked.

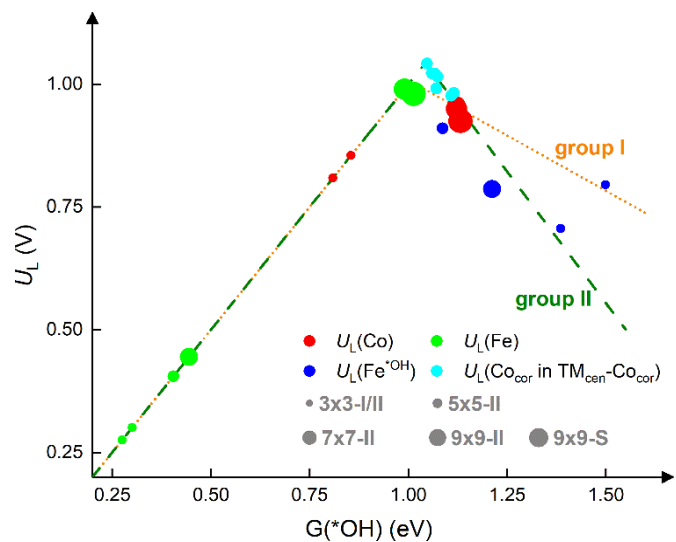

**Figure 9**. The calculated limiting potential ($U_L$) for ORR as a function of the Gibbs free energy of *OH ($G(*OH)$). The orange dotted line and the olive dashed line are the predicted limiting potential based on the linear relation fitted to the sites in group I and group II in **Figure 8b**, respectively.

**References**

1. R. Qin, K. Liu, Q. Wu and N. Zheng, *Chem Rev*, 2020, **120**, 11810-11899.
2. Y. Wang, H. Su, Y. He, L. Li, S. Zhu, H. Shen, P. Xie, X. Fu, G. Zhou, C. Feng, D. Zhao, F. Xiao, X. Zhu, Y. Zeng, M. Shao, S. Chen, G. Wu, J. Zeng and C. Wang, *Chem. Rev.*, 2020, DOI: 10.1021/acs.chemrev.0c00594.
3. Y. Wang, D. Wang and Y. Li, *Adv. Mater.*, 2021, **33**, 2008151.
4. W. Zang, T. Sun, T. Yang, S. Xi, M. Waqar, Z. Kou, Z. Lyu, Y. P. Feng, J. Wang and S. J. Pennycook, *Adv Mater*, 2021, **33**, e2003846.
5. Y. Xue, B. Huang, Y. Yi, Y. Guo, Z. Zuo, Y. Li, Z. Jia, H. Liu and Y. Li, *Nat. Commun.*, 2018, **9**, 1460.
6. L. Cao, Q. Luo, W. Liu, Y. Lin, X. Liu, Y. Cao, W. Zhang, Y. Wu, J. Yang, T. Yao and S. Wei, *Nat. Catal.*, 2019, **2**, 134-141.
7. S. Fang, X. Zhu, X. Liu, J. Gu, W. Liu, D. Wang, W. Zhang, Y. Lin, J. Lu, S. Wei, Y. Li and T. Yao, *Nat. Commun.*, 2020, **11**, 1029.
8. Z. Zhang, C. Feng, C. Liu, M. Zuo, L. Qin, X. Yan, Y. Xing, H. Li, R. Si, S. Zhou and J. Zeng, *Nat Commun*, 2020, **11**, 1215.
9. Y. Shi, W.-M. Huang, J. Li, Y. Zhou, Z.-Q. Li, Y.-C. Yin and X.-H. Xia, *Nat. Commun.*, 2020, **11**, 4558.
10. H. T. Chung, D. A. Cullen, D. Higgins, B. T. Sneed, E. F. Holby, K. L. More and P. Zelenay, *Science*, 2017, **357**, 479-484.
11. X. X. Wang, D. A. Cullen, Y.-T. Pan, S. Hwang, M. Wang, Z. Feng, J. Wang, M. H. Engelhard, H. Zhang, Y. He, Y. Shao, D. Su, K. L. More, J. S. Spendelow and G. Wu, *Adv. Mater.*, 2018, **30**, 1706758.
12. H. Zhang, Y. Liu, T. Chen, J. Zhang, J. Zhang and X. W. D. Lou, *Adv Mater*, 2019, **31**, e1904548.
13. J. Liu, M. Jiao, L. Lu, H. M. Barkholtz, Y. Li, Y. Wang, L. Jiang, Z. Wu, D.-j. Liu, L. Zhuang, C. Ma, J. Zeng, B. Zhang, D. Su, P. Song, W. Xing, W. Xu, Y. Wang, Z. Jiang and G. Sun, *Nat. Commun.*, 2017, **8**, 15938.
14. H. Zhou, T. Yang, Z. Kou, L. Shen, Y. Zhao, Z. Wang, X. Wang, Z. Yang, J. Du, J. Xu, M. Chen, L. Tian, W. Guo, Q. Wang, H. Lv, W. Chen, X. Hong, J. Luo, D. He and Y. Wu, *Angew. Chem. Int. Ed.*, 2020, **59**, 20465-20469.
15. W. Zang, T. Yang, H. Zou, S. Xi, H. Zhang, X. Liu, Z. Kou, Y. Du, Y. P. Feng, L. Shen, L. Duan, J. Wang and S. J. Pennycook, *ACS Catal.*, 2019, **9**, 10166-10173.

16. T. Yang, T. T. Song, J. Zhou, S. Wang, D. Chi, L. Shen, M. Yang and Y. P. Feng, *Nano Energy*, 2020, **68**, 104304.
17. C. Ling, Y. Ouyang, Q. Li, X. Bai, X. Mao, A. Du and J. Wang, *Small Methods*, 2019, **3**, 1800376.
18. Z. Geng, Y. Liu, X. Kong, P. Li, K. Li, Z. Liu, J. Du, M. Shu, R. Si and J. Zeng, *Adv. Mater.*, 2018, **30**, 1803498.
19. W. Ju, A. Bagger, G.-P. Hao, A. S. Varela, I. Sinev, V. Bon, B. Roldan Cuenya, S. Kaskel, J. Rossmeisl and P. Strasser, *Nat. Commun.*, 2017, **8**, 944.
20. D. Xi, J. Li, J. Low, K. Mao, R. Long, J. Li, Z. Dai, T. Shao, Y. Zhong, Y. Li, Z. Li, X. J. Loh, L. Song, E. Ye and Y. Xiong, *Adv Mater*, 2021, **n/a**, e2104090.
21. H. Yin, P. Yuan, B.-A. Lu, H. Xia, K. Guo, G. Yang, G. Qu, D. Xue, Y. Hu, J. Cheng, S. Mu and J.-N. Zhang, *ACS Catal.*, 2021, **11**, 12754-12762.
22. R. Jiang, L. Li, T. Sheng, G. Hu, Y. Chen and L. Wang, *J. Am. Chem. Soc.*, 2018, **140**, 11594-11598.
23. A. Han, X. Wang, K. Tang, Z. Zhang, C. Ye, K. Kong, H. Hu, L. Zheng, P. Jiang, C. Zhao, Q. Zhang, D. Wang and Y. Li, *Angew. Chem. Int. Ed.*, 2021, **60**, 19262-19271.
24. C. Tang, L. Chen, H. Li, L. Li, Y. Jiao, Y. Zheng, H. Xu, K. Davey and S.-Z. Qiao, *J. Am. Chem. Soc.*, 2021, **143**, 7819-7827.
25. E. Jung, H. Shin, B.-H. Lee, V. Efremov, S. Lee, H. S. Lee, J. Kim, W. Hooch Antink, S. Park, K.-S. Lee, S.-P. Cho, J. S. Yoo, Y.-E. Sung and T. Hyeon, *Nat. Mater.*, 2020, **19**, 436-442.
26. J. Zhang, Y. Zhao, C. Chen, Y.-C. Huang, C.-L. Dong, C.-J. Chen, R.-S. Liu, C. Wang, K. Yan, Y. Li and G. Wang, *J. Am. Chem. Soc.*, 2019, **141**, 20118-20126.
27. K. Jiang, S. Back, A. J. Akey, C. Xia, Y. Hu, W. Liang, D. Schaak, E. Stavitski, J. K. Nørskov, S. Siahrostami and H. Wang, *Nat. Commun.*, 2019, **10**, 3997.
28. X. Hai, X. Zhao, N. Guo, C. Yao, C. Chen, W. Liu, Y. Du, H. Yan, J. Li, Z. Chen, X. Li, Z. Li, H. Xu, P. Lyu, J. Zhang, M. Lin, C. Su, S. J. Pennycook, C. Zhang, S. Xi and J. Lu, *ACS Catal.*, 2020, **10**, 5862-5870.
29. D. Malko, A. Kucernak and T. Lopes, *Nat. Commun.*, 2016, **7**, 13285.
30. X. Wan, X. Liu, Y. Li, R. Yu, L. Zheng, W. Yan, H. Wang, M. Xu and J. Shui, *Nat. Catal.*, 2019, **2**, 259-268.
31. Q.-K. Li, X.-F. Li, G. Zhang and J. Jiang, *J. Am. Chem. Soc.*, 2018, **140**, 15149-15152.
32. Y. Han, Q.-K. Li, K. Ye, Y. Luo, J. Jiang and G. Zhang, *ACS Appl. Mater. Interfaces*, 2020, **12**, 15271-15278.

33. Z. Jin, P. Li, Y. Meng, Z. Fang, D. Xiao and G. Yu, *Nat. Catal.*, 2021, **4**, 615-622.
34. X. Han, X. Ling, D. Yu, D. Xie, L. Li, S. Peng, C. Zhong, N. Zhao, Y. Deng and W. Hu, *Adv. Mater.*, 2019, **31**, 1905622.
35. M. A. Hunter, J. M. T. A. Fischer, Q. Yuan, M. Hankel and D. J. Searles, *ACS Catal.*, 2019, **9**, 7660-7667.
36. H. Li, Y. Wen, M. Jiang, Y. Yao, H. Zhou, Z. Huang, J. Li, S. Jiao, Y. Kuang and S. Luo, *Adv. Funct. Mater.*, 2021, **31**, 2011289.
37. J. Wang, W. Liu, G. Luo, Z. Li, C. Zhao, H. Zhang, M. Zhu, Q. Xu, X. Wang, C. Zhao, Y. Qu, Z. Yang, T. Yao, Y. Li, Y. Lin, Y. Wu and Y. Li, *Energy Environ. Sci.*, 2018, **11**, 3375-3379.
38. D. Zhang, W. Chen, Z. Li, Y. Chen, L. Zheng, Y. Gong, Q. Li, R. Shen, Y. Han, W.-C. Cheong, L. Gu and Y. Li, *Chem. Commun.*, 2018, **54**, 4274-4277.
39. B. Hu, A. Huang, X. Zhang, Z. Chen, R. Tu, W. Zhu, Z. Zhuang, C. Chen, Q. Peng and Y. Li, *Nano Res.*, 2021, DOI: 10.1007/s12274-021-3535-4.
40. X. Hai, S. Xi, S. Mitchell, K. Harrath, H. Xu, D. F. Akl, D. Kong, J. Li, Z. Li, T. Sun, H. Yang, Y. Cui, C. Su, X. Zhao, J. Li, J. Pérez-Ramírez and J. Lu, *Nat. Nanotechnol.*, 2021, DOI: 10.1038/s41565-021-01022-y.
41. X.-F. Yang, A. Wang, B. Qiao, J. Li, J. Liu and T. Zhang, *Acc. Chem. Res.*, 2013, **46**, 1740-1748.
42. X. Guo, S. Lin, J. Gu, S. Zhang, Z. Chen and S. Huang, *ACS Catal.*, 2019, **9**, 11042-11054.
43. H. Peng, F. Liu, X. Liu, S. Liao, C. You, X. Tian, H. Nan, F. Luo, H. Song, Z. Fu and P. Huang, *ACS Catal.*, 2014, **4**, 3797-3805.
44. J. Wang, Z. Huang, W. Liu, C. Chang, H. Tang, Z. Li, W. Chen, C. Jia, T. Yao, S. Wei, Y. Wu and Y. Li, *J. Am. Chem. Soc.*, 2017, **139**, 17281-17284.
45. Y. Zhang, T. Hou, Q. Xu, Q. Wang, Y. Bai, S. Yang, D. Rao, L. Wu, H. Pan, J. Chen, G. Wang, J. Zhu, T. Yao and X. Zheng, *Adv. Sci.*, 2021, **8**, 2100302.
46. X.-F. Li, Q.-K. Li, J. Cheng, L. Liu, Q. Yan, Y. Wu, X.-H. Zhang, Z.-Y. Wang, Q. Qiu and Y. Luo, *J. Am. Chem. Soc.*, 2016, **138**, 8706-8709.
47. J. Hwang, R. R. Rao, L. Giordano, Y. Katayama, Y. Yu and Y. Shao-Horn, *Science*, 2017, **358**, 751-756.
48. F. Li, H. Ai, D. Liu, K. H. Lo and H. Pan, *J. Mater. Chem. A*, 2021, **9**, 17749-17759.
49. U. I. Kramm, M. Lefèvre, N. Larouche, D. Schmeisser and J.-P. Dodelet, *J. Am. Chem. Soc.*, 2014, **136**, 978-985.

50. X. Li, C.-S. Cao, S.-F. Hung, Y.-R. Lu, W. Cai, A. I. Rykov, S. Miao, S. Xi, H. Yang, Z. Hu, J. Wang, J. Zhao, E. E. Alp, W. Xu, T.-S. Chan, H. Chen, Q. Xiong, H. Xiao, Y. Huang, J. Li, T. Zhang and B. Liu, *Chem*, 2020, **6**, 3440-3454.
51. W. Zhong, Y. Qiu, H. Shen, X. Wang, J. Yuan, C. Jia, S. Bi and J. Jiang, *J Am Chem Soc*, 2021, **143**, 4405-4413.
52. G. Yang, J. Zhu, P. Yuan, Y. Hu, G. Qu, B.-A. Lu, X. Xue, H. Yin, W. Cheng, J. Cheng, W. Xu, J. Li, J. Hu, S. Mu and J.-N. Zhang, *Nat. Commun.*, 2021, **12**, 1734.
53. F. Li, G.-F. Han, H.-J. Noh, S.-J. Kim, Y. Lu, H. Y. Jeong, Z. Fu and J.-B. Baek, *Energy Environ. Sci.*, 2018, **11**, 2263-2269.
54. M. Xiao, J. Zhu, G. Li, N. Li, S. Li, Z. P. Cano, L. Ma, P. Cui, P. Xu, G. Jiang, H. Jin, S. Wang, T. Wu, J. Lu, A. Yu, D. Su and Z. Chen, *Angew. Chem. Int. Ed.*, 2019, **58**, 9640-9645.
55. H. Li, S. Kelly, D. Guevarra, Z. Wang, Y. Wang, J. A. Haber, M. Anand, G. T. K. K. Gunasooriya, C. S. Abraham, S. Vijay, J. M. Gregoire and J. K. Nørskov, *Nat. Catal.*, 2021, **4**, 463-468.
56. A. Kulkarni, S. Siahrostami, A. Patel and J. K. Nørskov, *Chem. Rev.*, 2018, **118**, 2302-2312.
57. H. Xu, D. Cheng, D. Cao and X. C. Zeng, *Nat. Catal.*, 2018, **1**, 339-348.
58. K. Ye, M. Hu, Q.-K. Li, Y. Luo, J. Jiang and G. Zhang, *J. Phys. Chem. Lett.*, 2021, **12**, 5233-5240.